\documentclass[12pt]{article}
\usepackage{amsmath}
\usepackage{amssymb}
\usepackage{epsfig}
\usepackage{cite}

\usepackage{epic}
\usepackage{fancybox}

\def\alf1{ {\alpha\over\pi} }

\begin{document}
\input{feynman} 
\begin{titlepage}
 
\begin{center}
{\Large Recoil Phase Effect in Exclusive B Decays:\\ 
Implications for CP Violation$^{\dagger}$
}
\end{center}

\vspace{2mm}
\begin{center}
  {\bf   B.F.L. Ward$^{a,b,c}$}

\vspace{2mm}
{\em $^a$Department of Physics and Astronomy,\\
  The University of Tennessee, Knoxville, Tennessee 37996-1200, USA,}\\
{\em $^b$SLAC, Stanford University, Stanford, California 94309, USA}\\
{\em $^c$CERN, Theory Division, CH-1211 Geneva 23, Switzerland.}
\end{center}


\vspace{5mm}
\begin{center}
{\bf   Abstract}
\end{center}

In the perturbative QCD approach to exclusive B decays to two light
mesons, the leading twist contribution corresponds to those diagrams
in the Lepage-Brodsky expansion in which the would be spectator
quark receives its recoil momentum via one gluon exchange. We show that
the resulting amplitude, which in the spectator model is real,
acquires an imaginary part which may be comparable in size to its real part.
Thus, this source of the strong interaction phase in the amplitude
must be taken into account in general to discuss, reliably, the expectations
for CP violation in B decays at any B-Factory type scenario. 
\vspace{10mm}
\begin{center}
{\it Presented at the 2001 Cracow Epiphany Conference}
\end{center}
\vspace{10mm}
\renewcommand{\baselinestretch}{0.1}
\footnoterule
\noindent
{\footnotesize
\begin{itemize}
\item[${\dagger}$]
Work partly supported 
by the US Department of Energy Contracts  DE-FG05-91ER40627
and   DE-AC03-76ER00515.
\end{itemize}
}

\end{titlepage}

With the start up of the SLAC and KEK and HERA-B B-Factories and with the
imminent upgrades the CESR and Tevatron machines to CP violation
in B decays capability, comes the need to clarify the theoretical
expectations for this phenomenon. One important aspect of this phenomenon
is the possible interplay between the strong and weak phases in the respective
decay amplitudes. In particular, in decays such as $B\rightarrow \pi\pi$,
where amplitudes with both tree level and penguin contributions
are involved, it is necessary to know all sources of
a possible difference in their strong phases as well as their weak
phases. In this communication, we point-out an important source
of a difference in the strong phases of penguins and tree contributions
that is generally overlooked in the literature~\cite{others1,others2}.
In Refs.~\cite{semdcy,dppkk,pipi,rks1}, we have always treated 
this new strong phase source rigorously. As we illustrate below, 
unless the particular CP asymmetry parameter manifests 
itself already with amplitudes
that only involve a single strong phase, this new strong phase
must be taken into account to get reliable theoretical control of the
respective parameter.\par
More precisely, the situation can already be seen in the diagrams
in Fig. 1 for the process $\bar B_s\rightarrow \rho K_S$,
which are to be evaluated in the perturbative QCD formalism
of Lepage and Brodsky in Ref.~\cite{L-B} following 
the development of Ref.~\cite{shbrod}. See also Ref.~\cite{simwylr}
for further applications of the methods in  Ref.~\cite{shbrod}.
\begin{figure}
\begin{center}
\epsfig{file=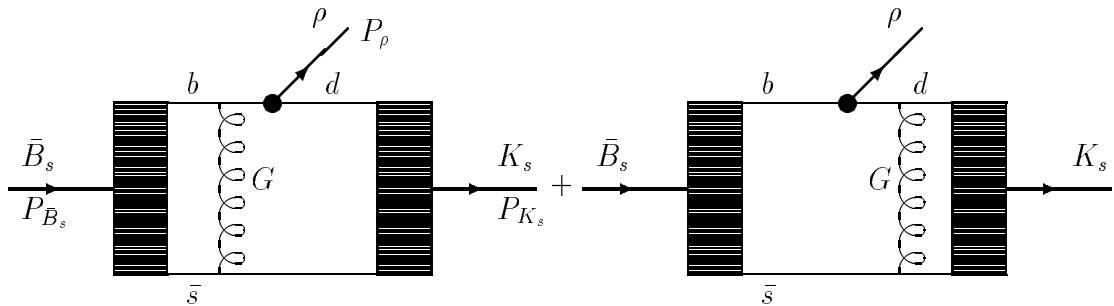}
\end{center}
\vspace{ -1.00cm}
\caption{\baselineskip=7mm     The process $\bar B_s
           \rightarrow           \rho
                 +       K_s$.
The four--momenta are indicated in the standard manner: $P_A$ is the
four--momentum of $A$ for all $A$.To leading order in the
perturbative QCD expansion defined by Lepage and Brodsky in
Ref.~\cite{L-B}, the two graphs shown are
the only ones that contribute in the factorisation ansatz
when penguins and colour exchange between the outgoing $\rho$
partons and the outgoing $K_s$ partons are ignored. The
remaining graphs in which the gluon $G$ is exchanged between
the would-be spectator $\bar s$ and the remaining $\rho$ parton lines
as well as the penguin type graphs are shown in Figs. 2 and 3,
where we see that, for QCD penguins, there is the added possibility 
that the gluon $G$ interacts with the penguin gluon itself of course.
}
\label{figone}
\end{figure}
The graph in Fig. 1a has the important property that, because
$m_B > m_b+m_s$, it is possible for the (heavy) b quark propagator
to reach its perturbative QCD mass shell. This generates an imaginary
part for this graph in comparison to the graph in Fig. 1b.
Similar conclusions hold for the graphs in Figs. 2 and 3 as 
\begin{figure}
\begin{center}
\epsfig{file=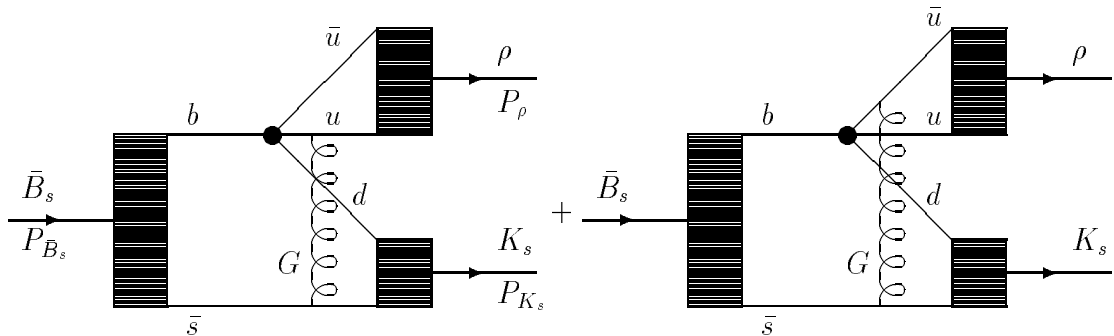}
\end{center}
\vspace{ -1.00cm}
\caption{\baselineskip=7mm     The colour exchange
graphs for the process $\bar B_s
           \rightarrow           \rho
                 +       K_s$ to leading order in the
Lepage-Brodsky expansion in Ref.~\cite{L-B}, ignoring penguins.
The kinematics is as defined in Fig.~\ref{figone}.
}
\label{figtwo}
\end{figure}
\begin{figure}
\begin{center}
\epsfig{file=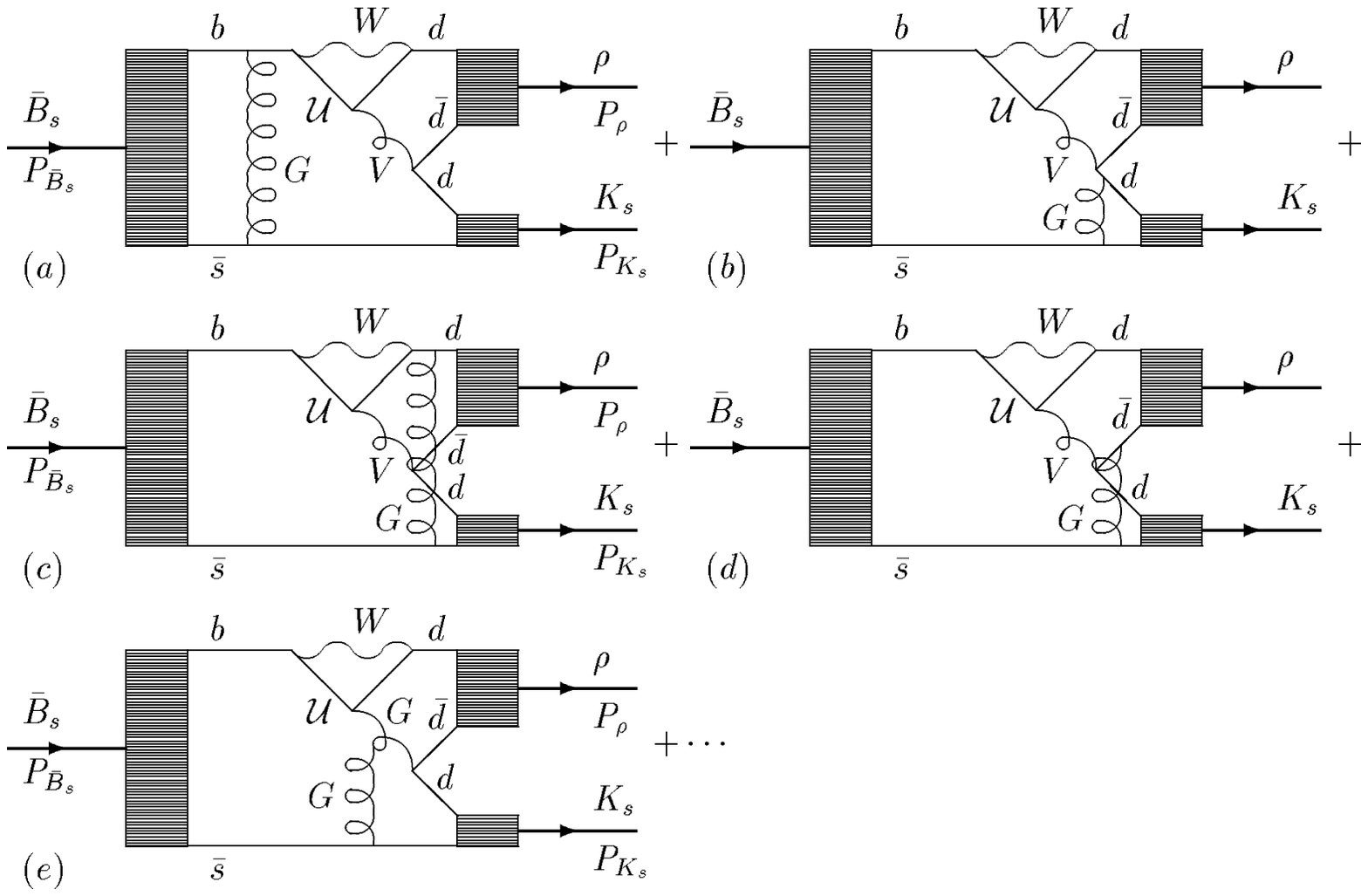}
\end{center}
\vspace{ -1.00cm}
\caption{\baselineskip=7mm     The penguin graphs for the process $\bar B_s
           \rightarrow           \rho
                 +       K_s$, to leading order in the
Lepage-Brodsky expansion defined in Ref.~\cite{L-B}.
The kinematics is as defined in Fig.~\ref{figone}.
}
\label{figthree}
\end{figure}
\noindent
well -- the graphs in which the would-be spectator 
receives its recoil 4-momentum
from the heavy quark line acquire an imaginary part. We refer to this
effect as the recoil phase effect~\cite{semdcy,dppkk,pipi,rks1,ccarlsn}.
This effect was always treated properly in our analyses in 
Refs.~\cite{semdcy,dppkk,pipi,rks1}. In Ref.~\cite{ccarlsn}, it was also 
treated properly. In Refs.~\cite{others1,others2}, 
it is not taken into account. In a recent 
analysis of the process $B\rightarrow \pi\pi$ in Ref.~\cite{beneke},
the dominant `Tree' recoil phase in the analogue of Fig. 1 is neglected
whereas the recoil phase in the diagrams in Fig. 2 and 3 are treated in some
approximation. Thus, the issue is quantitative. Does it really matter whether
one treats this recoil phase effect or not?\par

To answer this question, we use the results~\cite{rks1} 
we have obtained for the
process in Figs.~1-3. Specifically, we compute the decay width
$\Gamma(\bar B_s\rightarrow \rho K_S)$ and the penguin shift
of the CP violating angle $\gamma$'s sine, $\sin\gamma$,
where $\gamma$ is defined as in Ref.~\cite{sbfacwkp}. Here, following
Ref.~\cite{gronau}, we define the respective shift as $\Delta\sin\gamma$
which is given by
\begin{equation}
-\sin(2\gamma)-\Delta(\sin(2\gamma))\equiv {\Im \Lambda \over {1\over2}(1+|\Lambda|^2)} 
\label{pshift}
\end{equation}
for
\begin{equation}
\Lambda={A_Te^{-i\phi_T+i\delta_T}+\sum_j A_{P_j}e^{-i\phi_{P_j}+i\delta_{P_j}}
\over A_Te^{+i\phi_T+i\delta_T}+\sum_j
A_{P_j}e^{+i\phi_{P_j}+i\delta_{P_j}}},
\label{lambda}
\end{equation} 
where the amplitude $A_Te^{-i\phi_T+i\delta_T}$ corresponds to the 
tree-level weak processes
in Figs. 1 and 2  and the amplitudes 
$ A_{P_j}e^{-i\phi_{P_j}+i\delta_{P_j}}$ correspond to the respective penguin
processes in Fig. 3. Here, we identify the weak phases of the respective
amplitudes as $\phi_r$,~$r=T,P_j$ and the attendant strong phases as
$\delta_r$,~$r=T,P_j$. In general, $j=1,2$ distinguishes the electric
and magnetic penguins when this is required, as one can see in the
Appendix in Ref.~\cite{rks1}. In this notation, we have $\gamma\equiv \phi_T$.
\par
The details of our calculation are given in Ref.~\cite{rks1}. Here, for 
completeness, we summarise the basic theoretical framework.
Concerning the Cabibbo-Kobayashi-Maskawa (CKM) matrix itself,
we follow the conventions of Gilman and Kleinknecht
in Ref.~\cite{G-K} for the
CP-violating phase $\delta_{13} \equiv \delta$ and in view of the
current limits on it we consider the entire range 
$0\leq \delta \leq 2\pi$.  For the CKM matrix parameters $V_{td}$ and
$V_{ub}$ we also consider their extremal values from Ref.~\cite{G-K}
(the Particle Data Group (PDG) compilation).
To parametrise these
extremes, we use the notation defined in Ref.~\cite{rfleisch}
for $|V_{ub}/V_{cb}|$
in terms of the parameter $R_b= .385\pm .166$~\cite{G-K}.
All other CKM matrix 
element parameters are taken at their central values~\cite{G-K}. 
We note that the QCD corrections to the weak
interaction Lagrangian will be represented via
the QCD corrected effective
weak interaction Hamiltonian ${\cal H}_{eff}$
as it is defined in Ref.~\cite{rfleisch}
\begin{equation}
{\cal H}_{eff} = {G_F\over\sqrt{2}}\left[\sum_{j=u,c}{V^*}_{jq}V_{jb}
                  \left\{\sum^{2}_{k=1}Q^{jq}_k\tilde C_k(\mu)+
                   \sum^{10}_{k=3}Q^{q}_k\tilde C_k(\mu)\right\}\right]+h.c.
\end{equation}
where the Wilson coefficients $\tilde C_i$ and operators $Q_k$ are as given
in Ref.~\cite{rfleisch}, $G_F$ is Fermi's constant, $\mu$ is
is the renormalization scale and is of ${\cal O}(m_b)$ and here $q=s$.
The application of this effective weak interaction Hamiltonian
to our process $\bar B_s\rightarrow \rho K_S$ then proceeds according to
the realization of the Lepage-Brodsky expansion as
described in Ref.~\cite{shbrod}. This leads to the ``dominant''
contribution in which the $\rho$ is interpolated into the 
operator $O_2=Q_1$ in ${\cal H}_{eff}$ via the factorised
current matrix element $<\rho|\bar u(0)\gamma_\mu P_L u(0)|0>$,
~$P_L\equiv\frac{1}{2}(1-\gamma_5)$
so that the
respective remaining current in $O_2=Q_1$ is responsible for the $\bar B_s$ to
$K_S$ transition shown in Fig. 1, to which we refer as
the no colour exchange ``Tree'' contribution ($NC_T$).
In Fig.~\ref{figtwo}, we show the graphs in which colour is exchanged
between the would-be spectator $\bar s$ in Fig.~\ref{figone} and the
outgoing $\rho$ parton lines and in Fig.~\ref{figthree} we show
the respective penguin graphs: the dominant graphs according to
the prescription in Ref.~\cite{shbrod}
(3a,3b), the colour exchange graphs (3c,3d), and the exchange of the hard
gluon $G$ between the would-be spectator $\bar s$ and the penguin gluon 
itself for QCD penguins, (3e), which we also will classify as colour exchange.
The complete amplitude for the
process under study here is given by the sum of the contribution of the
graphs in Fig.~\ref{figone} and those of the graphs in 
Figs.~\ref{figtwo} and \ref{figthree},
to leading order in the Lepage-Brodsky expansion defined in Ref.~\cite{L-B}
and realized according to the prescription in Ref.~\cite{shbrod}
as we have just described for Fig. 1, for example. The complete result for the 
amplitude for $\bar B_s \rightarrow \rho +  K_S$ is given
in Ref.~\cite{rks1}, where its implications for the
measurement of the unitarity triangle angle $\gamma$ are
presented. Here, we investigate the recoil phase effect in this amplitude
in its various aspects from Figs. 1, 2, and 3 separately.
\par
We take for definiteness the central CKM values.
As the 
individual phases
which we present are purely due to strong interactions, 
we may proceed in this way without loss of physical information.
We also set the value of the effective weak interaction parameter
(here, note $\tilde C_1=C_2,~\tilde C_2=C_1$)
$a_2$, which is $C_2(m_B)+C_1(m_B)/N_c$ in perturbative QCD, to be
the recent phenomenological value $a_2\cong 0.24$
as found in Ref.~\cite{browder}, but, as it scales the weak interaction,
it will not affect the individual strong phases which we study.
When we combine the various contributions from Figs. 1-3
to form the entire amplitude, then the weak
parameters are important in determining the total phase variation of the
amplitude and its attendant CP violating properties, as we shall see.  
More precisely, we first isolate the recoil phase of the contribution
to the amplitude from the graphs in Fig.~1 . From our 
formulas in Ref.~\cite{rks1} we get the strong recoil phase
(all phases are in radians unless explicitly indicated otherwise)
\begin{equation}
\label{eq:tr-ph}
\delta_{NC_T} = 0.528 .
\end{equation}
Already, this is an important result, as it and its analoga have been missed
by all previous analyses of exclusive B and D decays to two light mesons
except the authors' analyses~\cite{semdcy,dppkk,pipi,rks1} and the analysis
in Ref.~\cite{ccarlsn}. Evidently, analyses such as that in 
Ref.~\cite{th-2000-101} which sometimes assume that $\delta_{NC_T}$ is zero
are misguided and incorrect. As we have checked 
following the procedures in Ref.~\cite{rks1},
variation of the fundamental parameters in our calculation does not change
the strong phases of our amplitude by more than $\sim 15\%$, so that the
result in (\ref{eq:tr-ph}) and its analoga in similar
B decays must be taken into account in CP violation studies.
\par
Continuing in this way, we compute the strong recoil phase effect 
for the graphs in Fig.~2 as
\begin{equation}
\label{eq:trce-ph}
\delta_{CE_T} = 0.295 ,
\end{equation}
where we use the notation introduced in Ref.~\cite{rks1}
to denote contribution from Fig.~2 as the colour exchange tree
contribution $CE_T$. Similarly, the graphs in Fig. 3(a,b,c,d)
have the strong recoil phases
\begin{equation}
\label{eq:pnc-ph}
\delta_{P_1} = 0.471,~\delta_{P_2} = 0.360 ,
\end{equation}
where $P_j$ denotes the electric($j=1$) or magnetic($j=2$)
penguin contribution respectively. The graphs in Figs. 3(c,d)
have the strong recoil phase
\begin{equation}
\label{eq:pce-ph}
\delta_{CE_P} = -0.318 ,
\end{equation}
where $CE_P$ denotes penguin graphs with colour exchange between
the quarks, so that the graph in Fig. 3e, which involves the
colour exchange between the the quarks in the $\bar B_s$ and $K_S$ mesons
and the penguin gluon, has the strong recoil phase $\delta_{CEG_P}$ 
which we calculate to be
\begin{equation}
\label{eq:pceg-ph}
\delta_{CEG_P} = 2.33,
\end{equation}
where we neglect the magnetic form factor in these last two results.
One comment is immediate: the different values of the 
strong recoil phases we find mean that they can not be ignored as some 
irrelevant over-all factor in either calculating the rates for the
exclusive B-decays or calculating the CP asymmetries in these decays.
\par
Indeed, if we set the phases in Eqs.(\ref{eq:tr-ph} - \ref{eq:pceg-ph})
to zero, we get a different set of results for the rate for the
decay and its penguin pollution of the time dependent asymmetry:
we find the total decay rate $\Gamma(\bar B_s\rightarrow \rho K_S)$
that satisfies 
\begin{equation}
\label{eq:dcyrate}
0.221\times 10^{-20}\text{GeV}((\frac{f_{B_s}}{0.141\text{GeV}})^2 \ge \Gamma(\bar B_s\rightarrow \rho K_S) \ge 0.160\times 10^{-20}\text{GeV}(\frac{f_{B_s}}{0.141\text{GeV}})^2
\end{equation}
and we find for example penguin shift of $\sin2\gamma$ plotted in Fig. 4.
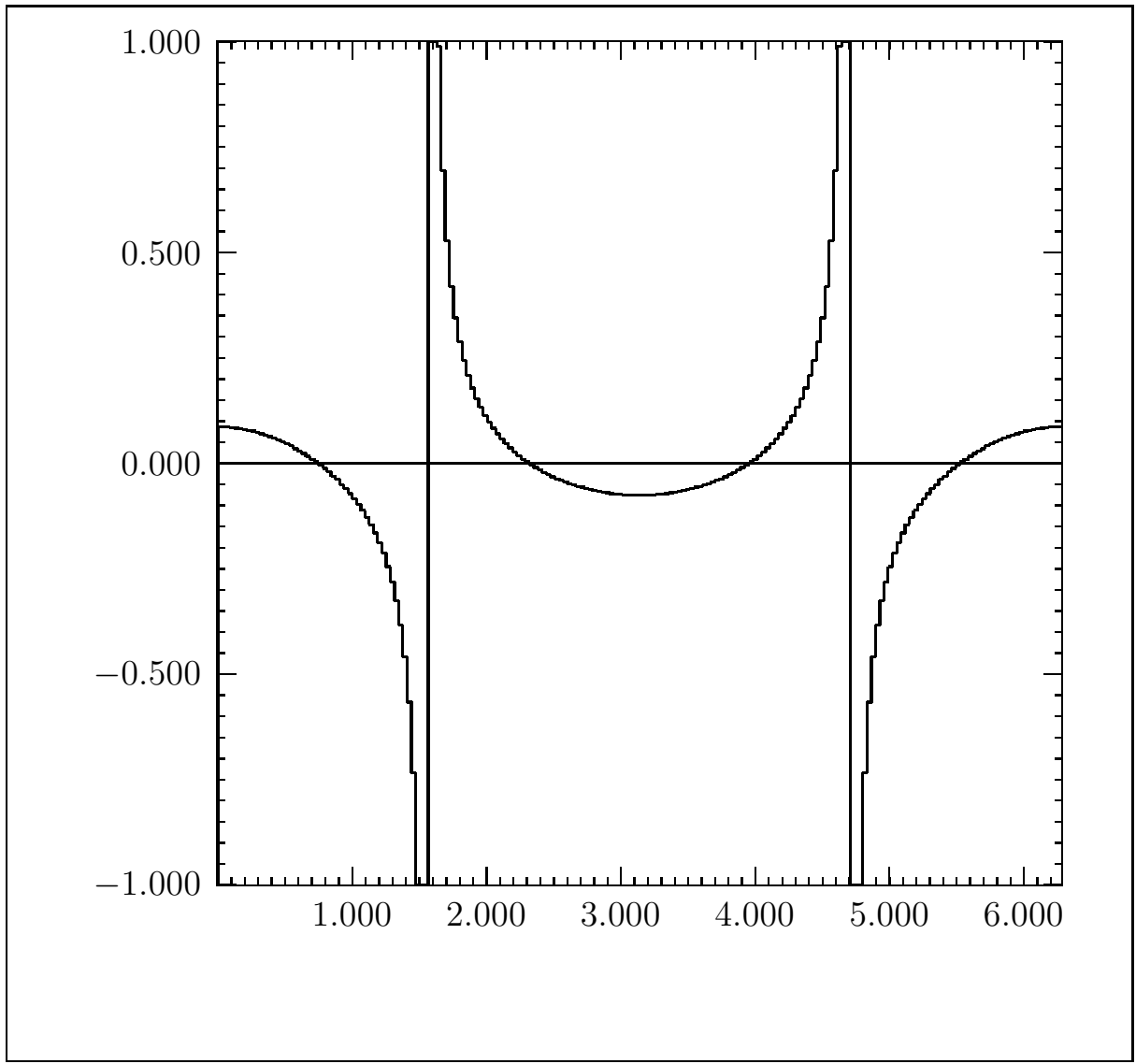
\begin{figure} 
\begin{center}
  Penguin Shift of $\sin(2\gamma)$ (Recoil Phase = 0 )                                               
\end{center}
\setlength{\unitlength}{0.1mm}
\begin{picture}(1600,1500)
\put(0,0){\framebox(1600,1500){ }}
\put(300,250){\begin{picture}( 1200,1200)
\put(0,0){\framebox( 1200,1200){ }}
\multiput(  190.99,0)(  190.99,0){   6}{\line(0,1){25}}
\multiput(     .00,0)(   19.10,0){  63}{\line(0,1){10}}
\multiput(  190.99,1200)(  190.99,0){   6}{\line(0,-1){25}}
\multiput(     .00,1200)(   19.10,0){  63}{\line(0,-1){10}}
\put( 190,-25){\makebox(0,0)[t]{\large $    1.000 $}}
\put( 381,-25){\makebox(0,0)[t]{\large $    2.000 $}}
\put( 572,-25){\makebox(0,0)[t]{\large $    3.000 $}}
\put( 763,-25){\makebox(0,0)[t]{\large $    4.000 $}}
\put( 954,-25){\makebox(0,0)[t]{\large $    5.000 $}}
\put(1145,-25){\makebox(0,0)[t]{\large $    6.000 $}}
\multiput(0,     .00)(0,  300.00){   5}{\line(1,0){25}}
\multiput(0,     .00)(0,   30.00){  41}{\line(1,0){10}}
\multiput(1200,     .00)(0,  300.00){   5}{\line(-1,0){25}}
\multiput(1200,     .00)(0,   30.00){  41}{\line(-1,0){10}}
\put(0,  600.00){\line(1,0){1200}}
\put(-25,   0){\makebox(0,0)[r]{\large $  -1.000 $}}
\put(-25, 300){\makebox(0,0)[r]{\large $   -0.500 $}}
\put(-25, 600){\makebox(0,0)[r]{\large $     0.000 $}}
\put(-25, 900){\makebox(0,0)[r]{\large $    0.500$}}
\put(-25,1200){\makebox(0,0)[r]{\large $   1.000 $}}
\end{picture}}
\put(300,250){\begin{picture}( 1200,1200)
\thicklines 
\newcommand{\x}[3]{\put(#1,#2){\line(1,0){#3}}}
\newcommand{\y}[3]{\put(#1,#2){\line(0,1){#3}}}
\newcommand{\z}[3]{\put(#1,#2){\line(0,-1){#3}}}
\newcommand{\e}[3]{\put(#1,#2){\line(0,1){#3}}}
\y{   0}{   0}{ 652}\x{   0}{ 652}{   5}
\y{   5}{ 652}{   0}\x{   5}{ 652}{   6}
\y{  11}{ 652}{   0}\x{  11}{ 652}{   6}
\z{  17}{ 652}{   1}\x{  17}{ 651}{   6}
\z{  23}{ 651}{   1}\x{  23}{ 650}{   6}
\z{  29}{ 650}{   1}\x{  29}{ 649}{   6}
\z{  35}{ 649}{   1}\x{  35}{ 648}{   6}
\z{  41}{ 648}{   1}\x{  41}{ 647}{   6}
\z{  47}{ 647}{   1}\x{  47}{ 646}{   6}
\z{  53}{ 646}{   2}\x{  53}{ 644}{   6}
\z{  59}{ 644}{   2}\x{  59}{ 642}{   6}
\z{  65}{ 642}{   2}\x{  65}{ 640}{   6}
\z{  71}{ 640}{   2}\x{  71}{ 638}{   6}
\z{  77}{ 638}{   2}\x{  77}{ 636}{   6}
\z{  83}{ 636}{   3}\x{  83}{ 633}{   6}
\z{  89}{ 633}{   2}\x{  89}{ 631}{   6}
\z{  95}{ 631}{   3}\x{  95}{ 628}{   6}
\z{ 101}{ 628}{   3}\x{ 101}{ 625}{   6}
\z{ 107}{ 625}{   4}\x{ 107}{ 621}{   6}
\z{ 113}{ 621}{   3}\x{ 113}{ 618}{   6}
\z{ 119}{ 618}{   4}\x{ 119}{ 614}{   6}
\z{ 125}{ 614}{   4}\x{ 125}{ 610}{   6}
\z{ 131}{ 610}{   4}\x{ 131}{ 606}{   6}
\z{ 137}{ 606}{   4}\x{ 137}{ 602}{   6}
\z{ 143}{ 602}{   5}\x{ 143}{ 597}{   6}
\z{ 149}{ 597}{   4}\x{ 149}{ 593}{   6}
\z{ 155}{ 593}{   5}\x{ 155}{ 588}{   6}
\z{ 161}{ 588}{   6}\x{ 161}{ 582}{   6}
\z{ 167}{ 582}{   5}\x{ 167}{ 577}{   6}
\z{ 173}{ 577}{   7}\x{ 173}{ 570}{   6}
\z{ 179}{ 570}{   6}\x{ 179}{ 564}{   6}
\z{ 185}{ 564}{   7}\x{ 185}{ 557}{   6}
\z{ 191}{ 557}{   7}\x{ 191}{ 550}{   6}
\z{ 197}{ 550}{   8}\x{ 197}{ 542}{   6}
\z{ 203}{ 542}{   9}\x{ 203}{ 533}{   6}
\z{ 209}{ 533}{  10}\x{ 209}{ 523}{   6}
\z{ 215}{ 523}{  10}\x{ 215}{ 513}{   6}
\z{ 221}{ 513}{  12}\x{ 221}{ 501}{   6}
\z{ 227}{ 501}{  14}\x{ 227}{ 487}{   6}
\z{ 233}{ 487}{  15}\x{ 233}{ 472}{   6}
\z{ 239}{ 472}{  19}\x{ 239}{ 453}{   6}
\z{ 245}{ 453}{  22}\x{ 245}{ 431}{   6}
\z{ 251}{ 431}{  26}\x{ 251}{ 405}{   6}
\z{ 257}{ 405}{  35}\x{ 257}{ 370}{   6}
\z{ 263}{ 370}{  45}\x{ 263}{ 325}{   6}
\z{ 269}{ 325}{  65}\x{ 269}{ 260}{   6}
\z{ 275}{ 260}{ 100}\x{ 275}{ 160}{   6}
\z{ 281}{ 160}{ 160}\x{ 281}{   0}{   6}
\y{ 287}{   0}{   0}\x{ 287}{   0}{   6}
\y{ 293}{   0}{   0}\x{ 293}{   0}{   6}
\y{ 299}{   0}{1200}\x{ 299}{1200}{   6}
\y{ 305}{1200}{   0}\x{ 305}{1200}{   6}
\z{ 311}{1200}{   6}\x{ 311}{1194}{   6}
\z{ 317}{1194}{ 178}\x{ 317}{1016}{   6}
\z{ 323}{1016}{  99}\x{ 323}{ 917}{   6}
\z{ 329}{ 917}{  65}\x{ 329}{ 852}{   6}
\z{ 335}{ 852}{  45}\x{ 335}{ 807}{   6}
\z{ 341}{ 807}{  34}\x{ 341}{ 773}{   6}
\z{ 347}{ 773}{  27}\x{ 347}{ 746}{   6}
\z{ 353}{ 746}{  21}\x{ 353}{ 725}{   6}
\z{ 359}{ 725}{  18}\x{ 359}{ 707}{   6}
\z{ 365}{ 707}{  15}\x{ 365}{ 692}{   6}
\z{ 371}{ 692}{  12}\x{ 371}{ 680}{   6}
\z{ 377}{ 680}{  12}\x{ 377}{ 668}{   6}
\z{ 383}{ 668}{   9}\x{ 383}{ 659}{   6}
\z{ 389}{ 659}{   9}\x{ 389}{ 650}{   6}
\z{ 395}{ 650}{   8}\x{ 395}{ 642}{   6}
\z{ 401}{ 642}{   7}\x{ 401}{ 635}{   6}
\z{ 407}{ 635}{   7}\x{ 407}{ 628}{   6}
\z{ 413}{ 628}{   6}\x{ 413}{ 622}{   6}
\z{ 419}{ 622}{   6}\x{ 419}{ 616}{   6}
\z{ 425}{ 616}{   5}\x{ 425}{ 611}{   6}
\z{ 431}{ 611}{   5}\x{ 431}{ 606}{   6}
\z{ 437}{ 606}{   4}\x{ 437}{ 602}{   6}
\z{ 443}{ 602}{   4}\x{ 443}{ 598}{   6}
\z{ 449}{ 598}{   4}\x{ 449}{ 594}{   6}
\z{ 455}{ 594}{   3}\x{ 455}{ 591}{   6}
\z{ 461}{ 591}{   4}\x{ 461}{ 587}{   6}
\z{ 467}{ 587}{   3}\x{ 467}{ 584}{   6}
\z{ 473}{ 584}{   3}\x{ 473}{ 581}{   6}
\z{ 479}{ 581}{   3}\x{ 479}{ 578}{   6}
\z{ 485}{ 578}{   2}\x{ 485}{ 576}{   6}
\z{ 491}{ 576}{   2}\x{ 491}{ 574}{   6}
\z{ 497}{ 574}{   3}\x{ 497}{ 571}{   6}
\z{ 503}{ 571}{   2}\x{ 503}{ 569}{   6}
\z{ 509}{ 569}{   2}\x{ 509}{ 567}{   6}
\z{ 515}{ 567}{   1}\x{ 515}{ 566}{   6}
\z{ 521}{ 566}{   2}\x{ 521}{ 564}{   6}
\z{ 527}{ 564}{   1}\x{ 527}{ 563}{   6}
\z{ 533}{ 563}{   2}\x{ 533}{ 561}{   6}
\z{ 539}{ 561}{   1}\x{ 539}{ 560}{   6}
\z{ 545}{ 560}{   1}\x{ 545}{ 559}{   6}
\z{ 551}{ 559}{   1}\x{ 551}{ 558}{   6}
\z{ 557}{ 558}{   1}\x{ 557}{ 557}{   6}
\z{ 563}{ 557}{   1}\x{ 563}{ 556}{   6}
\y{ 569}{ 556}{   0}\x{ 569}{ 556}{   6}
\z{ 575}{ 556}{   1}\x{ 575}{ 555}{   6}
\y{ 581}{ 555}{   0}\x{ 581}{ 555}{   6}
\y{ 587}{ 555}{   0}\x{ 587}{ 555}{   6}
\y{ 593}{ 555}{   0}\x{ 593}{ 555}{   6}
\y{ 599}{ 555}{   0}\x{ 599}{ 555}{   6}
\y{ 605}{ 555}{   0}\x{ 605}{ 555}{   6}
\y{ 611}{ 555}{   0}\x{ 611}{ 555}{   6}
\y{ 617}{ 555}{   0}\x{ 617}{ 555}{   6}
\y{ 623}{ 555}{   1}\x{ 623}{ 556}{   6}
\y{ 629}{ 556}{   0}\x{ 629}{ 556}{   6}
\y{ 635}{ 556}{   1}\x{ 635}{ 557}{   6}
\y{ 641}{ 557}{   1}\x{ 641}{ 558}{   6}
\y{ 647}{ 558}{   1}\x{ 647}{ 559}{   6}
\y{ 653}{ 559}{   1}\x{ 653}{ 560}{   6}
\y{ 659}{ 560}{   1}\x{ 659}{ 561}{   6}
\y{ 665}{ 561}{   2}\x{ 665}{ 563}{   6}
\y{ 671}{ 563}{   1}\x{ 671}{ 564}{   6}
\y{ 677}{ 564}{   2}\x{ 677}{ 566}{   6}
\y{ 683}{ 566}{   1}\x{ 683}{ 567}{   6}
\y{ 689}{ 567}{   2}\x{ 689}{ 569}{   6}
\y{ 695}{ 569}{   2}\x{ 695}{ 571}{   6}
\y{ 701}{ 571}{   3}\x{ 701}{ 574}{   6}
\y{ 707}{ 574}{   2}\x{ 707}{ 576}{   6}
\y{ 713}{ 576}{   2}\x{ 713}{ 578}{   6}
\y{ 719}{ 578}{   3}\x{ 719}{ 581}{   6}
\y{ 725}{ 581}{   3}\x{ 725}{ 584}{   6}
\y{ 731}{ 584}{   3}\x{ 731}{ 587}{   6}
\y{ 737}{ 587}{   4}\x{ 737}{ 591}{   6}
\y{ 743}{ 591}{   3}\x{ 743}{ 594}{   6}
\y{ 749}{ 594}{   4}\x{ 749}{ 598}{   6}
\y{ 755}{ 598}{   4}\x{ 755}{ 602}{   6}
\y{ 761}{ 602}{   4}\x{ 761}{ 606}{   6}
\y{ 767}{ 606}{   5}\x{ 767}{ 611}{   6}
\y{ 773}{ 611}{   5}\x{ 773}{ 616}{   6}
\y{ 779}{ 616}{   6}\x{ 779}{ 622}{   6}
\y{ 785}{ 622}{   6}\x{ 785}{ 628}{   6}
\y{ 791}{ 628}{   7}\x{ 791}{ 635}{   6}
\y{ 797}{ 635}{   7}\x{ 797}{ 642}{   6}
\y{ 803}{ 642}{   8}\x{ 803}{ 650}{   6}
\y{ 809}{ 650}{   9}\x{ 809}{ 659}{   6}
\y{ 815}{ 659}{   9}\x{ 815}{ 668}{   6}
\y{ 821}{ 668}{  12}\x{ 821}{ 680}{   6}
\y{ 827}{ 680}{  12}\x{ 827}{ 692}{   6}
\y{ 833}{ 692}{  15}\x{ 833}{ 707}{   6}
\y{ 839}{ 707}{  18}\x{ 839}{ 725}{   6}
\y{ 845}{ 725}{  21}\x{ 845}{ 746}{   6}
\y{ 851}{ 746}{  27}\x{ 851}{ 773}{   6}
\y{ 857}{ 773}{  34}\x{ 857}{ 807}{   6}
\y{ 863}{ 807}{  45}\x{ 863}{ 852}{   6}
\y{ 869}{ 852}{  65}\x{ 869}{ 917}{   6}
\y{ 875}{ 917}{  99}\x{ 875}{1016}{   6}
\y{ 881}{1016}{ 178}\x{ 881}{1194}{   6}
\y{ 887}{1194}{   6}\x{ 887}{1200}{   6}
\y{ 893}{1200}{   0}\x{ 893}{1200}{   6}
\z{ 899}{1200}{1200}\x{ 899}{   0}{   6}
\y{ 905}{   0}{   0}\x{ 905}{   0}{   6}
\y{ 911}{   0}{   0}\x{ 911}{   0}{   6}
\y{ 917}{   0}{ 160}\x{ 917}{ 160}{   6}
\y{ 923}{ 160}{ 100}\x{ 923}{ 260}{   6}
\y{ 929}{ 260}{  65}\x{ 929}{ 325}{   6}
\y{ 935}{ 325}{  45}\x{ 935}{ 370}{   6}
\y{ 941}{ 370}{  35}\x{ 941}{ 405}{   6}
\y{ 947}{ 405}{  26}\x{ 947}{ 431}{   6}
\y{ 953}{ 431}{  22}\x{ 953}{ 453}{   6}
\y{ 959}{ 453}{  19}\x{ 959}{ 472}{   6}
\y{ 965}{ 472}{  15}\x{ 965}{ 487}{   6}
\y{ 971}{ 487}{  14}\x{ 971}{ 501}{   6}
\y{ 977}{ 501}{  12}\x{ 977}{ 513}{   6}
\y{ 983}{ 513}{  10}\x{ 983}{ 523}{   6}
\y{ 989}{ 523}{  10}\x{ 989}{ 533}{   6}
\y{ 995}{ 533}{   9}\x{ 995}{ 542}{   6}
\y{1001}{ 542}{   8}\x{1001}{ 550}{   6}
\y{1007}{ 550}{   7}\x{1007}{ 557}{   6}
\y{1013}{ 557}{   7}\x{1013}{ 564}{   6}
\y{1019}{ 564}{   6}\x{1019}{ 570}{   6}
\y{1025}{ 570}{   7}\x{1025}{ 577}{   6}
\y{1031}{ 577}{   5}\x{1031}{ 582}{   6}
\y{1037}{ 582}{   6}\x{1037}{ 588}{   6}
\y{1043}{ 588}{   5}\x{1043}{ 593}{   6}
\y{1049}{ 593}{   4}\x{1049}{ 597}{   6}
\y{1055}{ 597}{   5}\x{1055}{ 602}{   6}
\y{1061}{ 602}{   4}\x{1061}{ 606}{   6}
\y{1067}{ 606}{   4}\x{1067}{ 610}{   6}
\y{1073}{ 610}{   4}\x{1073}{ 614}{   6}
\y{1079}{ 614}{   4}\x{1079}{ 618}{   6}
\y{1085}{ 618}{   3}\x{1085}{ 621}{   6}
\y{1091}{ 621}{   4}\x{1091}{ 625}{   6}
\y{1097}{ 625}{   3}\x{1097}{ 628}{   6}
\y{1103}{ 628}{   3}\x{1103}{ 631}{   6}
\y{1109}{ 631}{   2}\x{1109}{ 633}{   6}
\y{1115}{ 633}{   3}\x{1115}{ 636}{   6}
\y{1121}{ 636}{   2}\x{1121}{ 638}{   6}
\y{1127}{ 638}{   2}\x{1127}{ 640}{   6}
\y{1133}{ 640}{   2}\x{1133}{ 642}{   6}
\y{1139}{ 642}{   2}\x{1139}{ 644}{   6}
\y{1145}{ 644}{   2}\x{1145}{ 646}{   6}
\y{1151}{ 646}{   1}\x{1151}{ 647}{   6}
\y{1157}{ 647}{   1}\x{1157}{ 648}{   6}
\y{1163}{ 648}{   1}\x{1163}{ 649}{   6}
\y{1169}{ 649}{   1}\x{1169}{ 650}{   6}
\y{1175}{ 650}{   1}\x{1175}{ 651}{   6}
\y{1181}{ 651}{   1}\x{1181}{ 652}{   6}
\y{1187}{ 652}{   0}\x{1187}{ 652}{   6}
\y{1193}{ 652}{   0}\x{1193}{ 652}{   6}
\end{picture}} 
\end{picture} 
\caption{ \baselineskip=7mm
Penguin shift of the CP asymmetry $\sin(2\gamma)$ in
$\bar B_s\rightarrow \rho K_s$ for $R_b=0.385$ for the matrix element
with the recoil phase set to zero by using the principle value prescription
in the diagrams in Figs. 1 - 3. The analogous plots obtain
for the $\pm1\sigma$ values of $R_b$ as discussed in the text.}
\end{figure}
Thus, the shift is less than 29\% ( allowing a $3\sigma$ measurement of
$\sin2\gamma$) for $0\le \gamma \le 75.1^o$ and $103.4^o \le \gamma \le 180^o$.
These results should compared with the analogous presented in Ref.~\cite{rks1}
, where we found, when the recoil strong phases are not set to zero, that
\begin{equation}
\label{eq:dcyrate-rc}
0.495\times 10^{-20}\text{GeV}(\frac{f_{B_s}}{0.141\text{GeV}})^2 \ge \Gamma(\bar B_s\rightarrow \rho K_S) \ge 0.329\times 10^{-21}\text{GeV}(\frac{f_{B_s}}{0.141\text{GeV}})^2
\end{equation}
and that the shift is less than 29\% 
for $0\le \gamma \le 40.5^o$ and $102.5^o \le \gamma \le 157.9^o$.
The differences in these two sets of results show that
the recoil phase effect can not be ignored in exclusive
B decays of the type discussed in this paper.
\par
This brings us to a comparison of our analysis with those presented
in Refs.~\cite{others1,others2,beneke}. To illustrate the size
of the recoil phase effect, we use the $\bar B\rightarrow\pi\pi$
process which we have already analysed in Ref.~\cite{pipi}
and which Beneke {\it et al.} have treated in Ref.~\cite{beneke}.
From our Eq.(5) in Ref.~\cite{pipi} we see that, if the recoil
phases are set to zero in defining the integrals over the
light-cone fractions in the analogue of the diagrams in Figs.
1-3 here for the $\pi^+\pi^-$ case, the decay rates given in Eq.(8)
of Ref.~\cite{pipi} are changed by as much as$\sim  90\%$.
Moreover, if as Beneke {\it et al.} do, we set the recoil phase of
the 'dominant' Tree contribution in the analogue of Fig. 1 here,
these decays rates are still changed by as much as $\sim 90\%$.
Thus, none of the treatments of the recoil phase in 
Refs.~\cite{others1,others2,beneke} is sufficient.\par
The situation is entirely similar to the $\rho K_S$ case discussed
above insofar as the time dependent CP violating asymmetry is concerned
-- neither the complete neglect of the recoil phase
in Refs.~\cite{others1,others2} nor the neglect of the
recoil phase of the dominant 'Tree' contribution from
the analogue of Fig.~1 here as in Ref.~\cite{beneke} gives the
proper result shown in Fig. 4 of Ref.~\cite{pipi} for
the dependence of the penguin pollution on $\delta_{13}$.
To see how big the respective distortion can be on the 
CP violating asymmetry itself, we plot in Fig.~ 5 the value of
the direct CP violating asymmetry~\cite{th-2000-101}, 
${\cal A}^{dir}_{CP}(\pi\pi)$,
for the $\bar B\rightarrow\pi^+\pi^-$ case as derived from
Eq.(5) in Ref.~\cite{pipi}. 
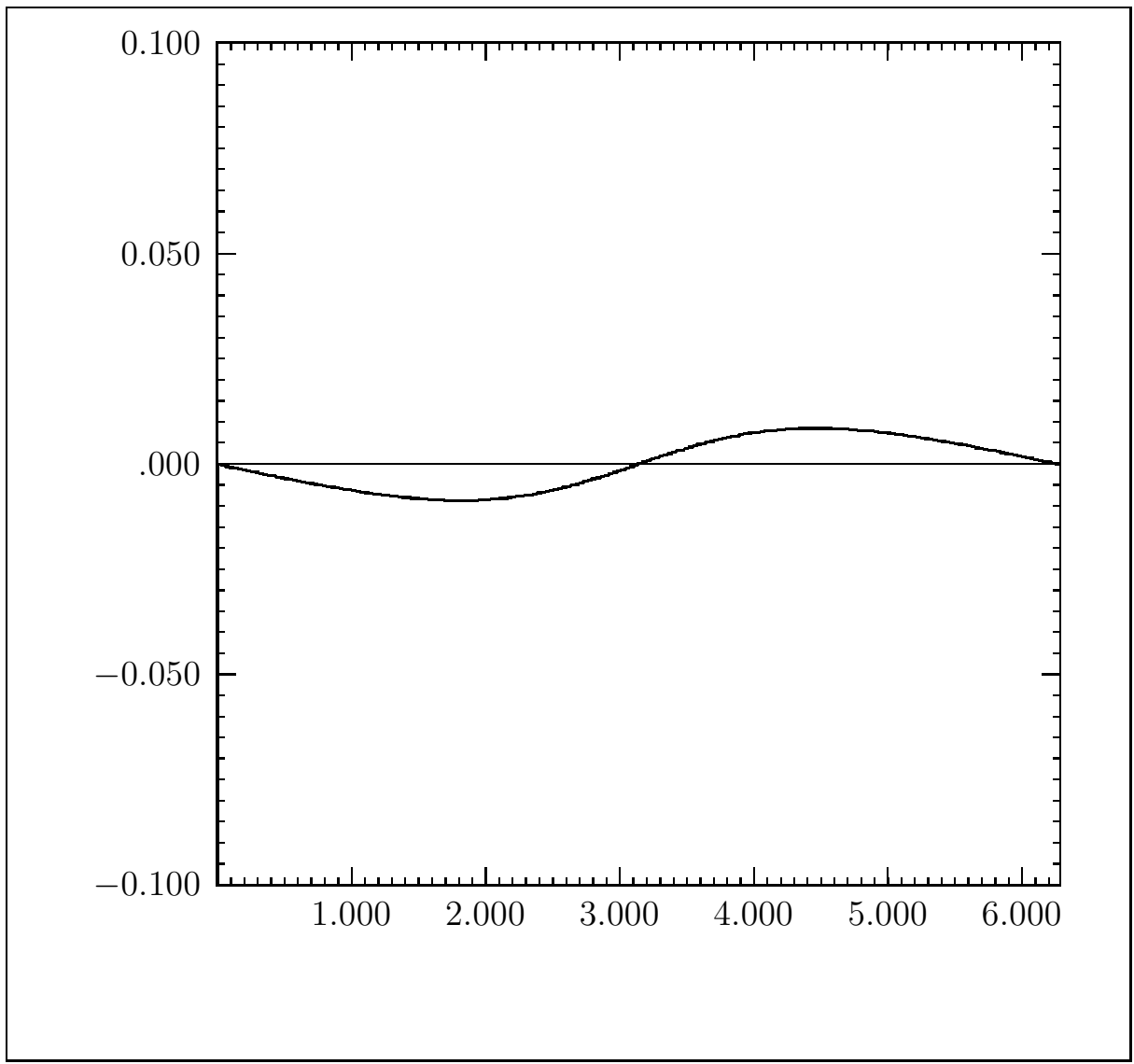
\begin{figure} 
\begin{center}
 Direct CP Asymmetry :${\cal A}^{dir}_{CP}(\pi\pi)_d$                                                  
\end{center}
\setlength{\unitlength}{0.1mm}
\begin{picture}(1600,1500)
\put(0,0){\framebox(1600,1500){ }}
\put(300,250){\begin{picture}( 1200,1200)
\put(0,0){\framebox( 1200,1200){ }}
\multiput(  190.99,0)(  190.99,0){   6}{\line(0,1){25}}
\multiput(     .00,0)(   19.10,0){  63}{\line(0,1){10}}
\multiput(  190.99,1200)(  190.99,0){   6}{\line(0,-1){25}}
\multiput(     .00,1200)(   19.10,0){  63}{\line(0,-1){10}}
\put( 190,-25){\makebox(0,0)[t]{\large $    1.000 $}}
\put( 381,-25){\makebox(0,0)[t]{\large $    2.000 $}}
\put( 572,-25){\makebox(0,0)[t]{\large $    3.000 $}}
\put( 763,-25){\makebox(0,0)[t]{\large $    4.000 $}}
\put( 954,-25){\makebox(0,0)[t]{\large $    5.000 $}}
\put(1145,-25){\makebox(0,0)[t]{\large $    6.000 $}}
\multiput(0,     .00)(0,  300.00){   5}{\line(1,0){25}}
\multiput(0,   30.00)(0,   30.00){  40}{\line(1,0){10}}
\multiput(1200,     .00)(0,  300.00){   5}{\line(-1,0){25}}
\multiput(1200,   30.00)(0,   30.00){  40}{\line(-1,0){10}}
\put(0,  600.00){\line(1,0){1200}}
\put(-25,   0){\makebox(0,0)[r]{\large $  -0.100 $}}
\put(-25, 299){\makebox(0,0)[r]{\large $   -0.050 $}}
\put(-25, 600){\makebox(0,0)[r]{\large $     .000 $}}
\put(-25, 900){\makebox(0,0)[r]{\large $    0.050 $}}
\put(-25,1200){\makebox(0,0)[r]{\large $   0.100 $}}
\end{picture}}
\put(300,250){\begin{picture}( 1200,1200)
\thicklines 
\newcommand{\x}[3]{\put(#1,#2){\line(1,0){#3}}}
\newcommand{\y}[3]{\put(#1,#2){\line(0,1){#3}}}
\newcommand{\z}[3]{\put(#1,#2){\line(0,-1){#3}}}
\newcommand{\e}[3]{\put(#1,#2){\line(0,1){#3}}}
\y{   0}{   0}{ 599}\x{   0}{ 599}{   5}
\z{   5}{ 599}{   1}\x{   5}{ 598}{   6}
\z{  11}{ 598}{   2}\x{  11}{ 596}{   6}
\z{  17}{ 596}{   1}\x{  17}{ 595}{   6}
\z{  23}{ 595}{   1}\x{  23}{ 594}{   6}
\z{  29}{ 594}{   1}\x{  29}{ 593}{   6}
\z{  35}{ 593}{   2}\x{  35}{ 591}{   6}
\z{  41}{ 591}{   1}\x{  41}{ 590}{   6}
\z{  47}{ 590}{   1}\x{  47}{ 589}{   6}
\z{  53}{ 589}{   1}\x{  53}{ 588}{   6}
\z{  59}{ 588}{   2}\x{  59}{ 586}{   6}
\z{  65}{ 586}{   1}\x{  65}{ 585}{   6}
\z{  71}{ 585}{   1}\x{  71}{ 584}{   6}
\z{  77}{ 584}{   1}\x{  77}{ 583}{   6}
\z{  83}{ 583}{   2}\x{  83}{ 581}{   6}
\z{  89}{ 581}{   1}\x{  89}{ 580}{   6}
\z{  95}{ 580}{   1}\x{  95}{ 579}{   6}
\z{ 101}{ 579}{   1}\x{ 101}{ 578}{   6}
\z{ 107}{ 578}{   1}\x{ 107}{ 577}{   6}
\z{ 113}{ 577}{   1}\x{ 113}{ 576}{   6}
\z{ 119}{ 576}{   2}\x{ 119}{ 574}{   6}
\z{ 125}{ 574}{   1}\x{ 125}{ 573}{   6}
\z{ 131}{ 573}{   1}\x{ 131}{ 572}{   6}
\z{ 137}{ 572}{   1}\x{ 137}{ 571}{   6}
\z{ 143}{ 571}{   1}\x{ 143}{ 570}{   6}
\z{ 149}{ 570}{   1}\x{ 149}{ 569}{   6}
\z{ 155}{ 569}{   1}\x{ 155}{ 568}{   6}
\z{ 161}{ 568}{   1}\x{ 161}{ 567}{   6}
\z{ 167}{ 567}{   1}\x{ 167}{ 566}{   6}
\z{ 173}{ 566}{   1}\x{ 173}{ 565}{   6}
\z{ 179}{ 565}{   1}\x{ 179}{ 564}{   6}
\z{ 185}{ 564}{   1}\x{ 185}{ 563}{   6}
\z{ 191}{ 563}{   1}\x{ 191}{ 562}{   6}
\z{ 197}{ 562}{   1}\x{ 197}{ 561}{   6}
\z{ 203}{ 561}{   1}\x{ 203}{ 560}{   6}
\z{ 209}{ 560}{   1}\x{ 209}{ 559}{   6}
\z{ 215}{ 559}{   1}\x{ 215}{ 558}{   6}
\z{ 221}{ 558}{   1}\x{ 221}{ 557}{   6}
\y{ 227}{ 557}{   0}\x{ 227}{ 557}{   6}
\z{ 233}{ 557}{   1}\x{ 233}{ 556}{   6}
\z{ 239}{ 556}{   1}\x{ 239}{ 555}{   6}
\z{ 245}{ 555}{   1}\x{ 245}{ 554}{   6}
\y{ 251}{ 554}{   0}\x{ 251}{ 554}{   6}
\z{ 257}{ 554}{   1}\x{ 257}{ 553}{   6}
\z{ 263}{ 553}{   1}\x{ 263}{ 552}{   6}
\y{ 269}{ 552}{   0}\x{ 269}{ 552}{   6}
\z{ 275}{ 552}{   1}\x{ 275}{ 551}{   6}
\y{ 281}{ 551}{   0}\x{ 281}{ 551}{   6}
\z{ 287}{ 551}{   1}\x{ 287}{ 550}{   6}
\y{ 293}{ 550}{   0}\x{ 293}{ 550}{   6}
\z{ 299}{ 550}{   1}\x{ 299}{ 549}{   6}
\y{ 305}{ 549}{   0}\x{ 305}{ 549}{   6}
\y{ 311}{ 549}{   0}\x{ 311}{ 549}{   6}
\y{ 317}{ 549}{   0}\x{ 317}{ 549}{   6}
\z{ 323}{ 549}{   1}\x{ 323}{ 548}{   6}
\y{ 329}{ 548}{   0}\x{ 329}{ 548}{   6}
\y{ 335}{ 548}{   0}\x{ 335}{ 548}{   6}
\y{ 341}{ 548}{   0}\x{ 341}{ 548}{   6}
\y{ 347}{ 548}{   0}\x{ 347}{ 548}{   6}
\y{ 353}{ 548}{   0}\x{ 353}{ 548}{   6}
\y{ 359}{ 548}{   0}\x{ 359}{ 548}{   6}
\y{ 365}{ 548}{   0}\x{ 365}{ 548}{   6}
\y{ 371}{ 548}{   1}\x{ 371}{ 549}{   6}
\y{ 377}{ 549}{   0}\x{ 377}{ 549}{   6}
\y{ 383}{ 549}{   0}\x{ 383}{ 549}{   6}
\y{ 389}{ 549}{   1}\x{ 389}{ 550}{   6}
\y{ 395}{ 550}{   0}\x{ 395}{ 550}{   6}
\y{ 401}{ 550}{   1}\x{ 401}{ 551}{   6}
\y{ 407}{ 551}{   0}\x{ 407}{ 551}{   6}
\y{ 413}{ 551}{   1}\x{ 413}{ 552}{   6}
\y{ 419}{ 552}{   1}\x{ 419}{ 553}{   6}
\y{ 425}{ 553}{   1}\x{ 425}{ 554}{   6}
\y{ 431}{ 554}{   0}\x{ 431}{ 554}{   6}
\y{ 437}{ 554}{   1}\x{ 437}{ 555}{   6}
\y{ 443}{ 555}{   1}\x{ 443}{ 556}{   6}
\y{ 449}{ 556}{   1}\x{ 449}{ 557}{   6}
\y{ 455}{ 557}{   2}\x{ 455}{ 559}{   6}
\y{ 461}{ 559}{   1}\x{ 461}{ 560}{   6}
\y{ 467}{ 560}{   1}\x{ 467}{ 561}{   6}
\y{ 473}{ 561}{   1}\x{ 473}{ 562}{   6}
\y{ 479}{ 562}{   2}\x{ 479}{ 564}{   6}
\y{ 485}{ 564}{   1}\x{ 485}{ 565}{   6}
\y{ 491}{ 565}{   2}\x{ 491}{ 567}{   6}
\y{ 497}{ 567}{   1}\x{ 497}{ 568}{   6}
\y{ 503}{ 568}{   2}\x{ 503}{ 570}{   6}
\y{ 509}{ 570}{   1}\x{ 509}{ 571}{   6}
\y{ 515}{ 571}{   2}\x{ 515}{ 573}{   6}
\y{ 521}{ 573}{   2}\x{ 521}{ 575}{   6}
\y{ 527}{ 575}{   2}\x{ 527}{ 577}{   6}
\y{ 533}{ 577}{   2}\x{ 533}{ 579}{   6}
\y{ 539}{ 579}{   1}\x{ 539}{ 580}{   6}
\y{ 545}{ 580}{   2}\x{ 545}{ 582}{   6}
\y{ 551}{ 582}{   2}\x{ 551}{ 584}{   6}
\y{ 557}{ 584}{   2}\x{ 557}{ 586}{   6}
\y{ 563}{ 586}{   2}\x{ 563}{ 588}{   6}
\y{ 569}{ 588}{   2}\x{ 569}{ 590}{   6}
\y{ 575}{ 590}{   2}\x{ 575}{ 592}{   6}
\y{ 581}{ 592}{   2}\x{ 581}{ 594}{   6}
\y{ 587}{ 594}{   2}\x{ 587}{ 596}{   6}
\y{ 593}{ 596}{   2}\x{ 593}{ 598}{   6}
\y{ 599}{ 598}{   3}\x{ 599}{ 601}{   6}
\y{ 605}{ 601}{   2}\x{ 605}{ 603}{   6}
\y{ 611}{ 603}{   2}\x{ 611}{ 605}{   6}
\y{ 617}{ 605}{   2}\x{ 617}{ 607}{   6}
\y{ 623}{ 607}{   2}\x{ 623}{ 609}{   6}
\y{ 629}{ 609}{   2}\x{ 629}{ 611}{   6}
\y{ 635}{ 611}{   2}\x{ 635}{ 613}{   6}
\y{ 641}{ 613}{   2}\x{ 641}{ 615}{   6}
\y{ 647}{ 615}{   2}\x{ 647}{ 617}{   6}
\y{ 653}{ 617}{   2}\x{ 653}{ 619}{   6}
\y{ 659}{ 619}{   1}\x{ 659}{ 620}{   6}
\y{ 665}{ 620}{   2}\x{ 665}{ 622}{   6}
\y{ 671}{ 622}{   2}\x{ 671}{ 624}{   6}
\y{ 677}{ 624}{   2}\x{ 677}{ 626}{   6}
\y{ 683}{ 626}{   2}\x{ 683}{ 628}{   6}
\y{ 689}{ 628}{   1}\x{ 689}{ 629}{   6}
\y{ 695}{ 629}{   2}\x{ 695}{ 631}{   6}
\y{ 701}{ 631}{   1}\x{ 701}{ 632}{   6}
\y{ 707}{ 632}{   2}\x{ 707}{ 634}{   6}
\y{ 713}{ 634}{   1}\x{ 713}{ 635}{   6}
\y{ 719}{ 635}{   2}\x{ 719}{ 637}{   6}
\y{ 725}{ 637}{   1}\x{ 725}{ 638}{   6}
\y{ 731}{ 638}{   1}\x{ 731}{ 639}{   6}
\y{ 737}{ 639}{   1}\x{ 737}{ 640}{   6}
\y{ 743}{ 640}{   2}\x{ 743}{ 642}{   6}
\y{ 749}{ 642}{   1}\x{ 749}{ 643}{   6}
\y{ 755}{ 643}{   1}\x{ 755}{ 644}{   6}
\y{ 761}{ 644}{   1}\x{ 761}{ 645}{   6}
\y{ 767}{ 645}{   0}\x{ 767}{ 645}{   6}
\y{ 773}{ 645}{   1}\x{ 773}{ 646}{   6}
\y{ 779}{ 646}{   1}\x{ 779}{ 647}{   6}
\y{ 785}{ 647}{   1}\x{ 785}{ 648}{   6}
\y{ 791}{ 648}{   0}\x{ 791}{ 648}{   6}
\y{ 797}{ 648}{   1}\x{ 797}{ 649}{   6}
\y{ 803}{ 649}{   0}\x{ 803}{ 649}{   6}
\y{ 809}{ 649}{   1}\x{ 809}{ 650}{   6}
\y{ 815}{ 650}{   0}\x{ 815}{ 650}{   6}
\y{ 821}{ 650}{   0}\x{ 821}{ 650}{   6}
\y{ 827}{ 650}{   1}\x{ 827}{ 651}{   6}
\y{ 833}{ 651}{   0}\x{ 833}{ 651}{   6}
\y{ 839}{ 651}{   0}\x{ 839}{ 651}{   6}
\y{ 845}{ 651}{   0}\x{ 845}{ 651}{   6}
\y{ 851}{ 651}{   0}\x{ 851}{ 651}{   6}
\y{ 857}{ 651}{   0}\x{ 857}{ 651}{   6}
\y{ 863}{ 651}{   0}\x{ 863}{ 651}{   6}
\y{ 869}{ 651}{   0}\x{ 869}{ 651}{   6}
\z{ 875}{ 651}{   1}\x{ 875}{ 650}{   6}
\y{ 881}{ 650}{   0}\x{ 881}{ 650}{   6}
\y{ 887}{ 650}{   0}\x{ 887}{ 650}{   6}
\y{ 893}{ 650}{   0}\x{ 893}{ 650}{   6}
\z{ 899}{ 650}{   1}\x{ 899}{ 649}{   6}
\y{ 905}{ 649}{   0}\x{ 905}{ 649}{   6}
\z{ 911}{ 649}{   1}\x{ 911}{ 648}{   6}
\y{ 917}{ 648}{   0}\x{ 917}{ 648}{   6}
\z{ 923}{ 648}{   1}\x{ 923}{ 647}{   6}
\y{ 929}{ 647}{   0}\x{ 929}{ 647}{   6}
\z{ 935}{ 647}{   1}\x{ 935}{ 646}{   6}
\z{ 941}{ 646}{   1}\x{ 941}{ 645}{   6}
\y{ 947}{ 645}{   0}\x{ 947}{ 645}{   6}
\z{ 953}{ 645}{   1}\x{ 953}{ 644}{   6}
\z{ 959}{ 644}{   1}\x{ 959}{ 643}{   6}
\z{ 965}{ 643}{   1}\x{ 965}{ 642}{   6}
\y{ 971}{ 642}{   0}\x{ 971}{ 642}{   6}
\z{ 977}{ 642}{   1}\x{ 977}{ 641}{   6}
\z{ 983}{ 641}{   1}\x{ 983}{ 640}{   6}
\z{ 989}{ 640}{   1}\x{ 989}{ 639}{   6}
\z{ 995}{ 639}{   1}\x{ 995}{ 638}{   6}
\z{1001}{ 638}{   1}\x{1001}{ 637}{   6}
\z{1007}{ 637}{   1}\x{1007}{ 636}{   6}
\z{1013}{ 636}{   1}\x{1013}{ 635}{   6}
\z{1019}{ 635}{   1}\x{1019}{ 634}{   6}
\z{1025}{ 634}{   1}\x{1025}{ 633}{   6}
\z{1031}{ 633}{   1}\x{1031}{ 632}{   6}
\z{1037}{ 632}{   1}\x{1037}{ 631}{   6}
\z{1043}{ 631}{   1}\x{1043}{ 630}{   6}
\z{1049}{ 630}{   1}\x{1049}{ 629}{   6}
\z{1055}{ 629}{   1}\x{1055}{ 628}{   6}
\z{1061}{ 628}{   1}\x{1061}{ 627}{   6}
\z{1067}{ 627}{   1}\x{1067}{ 626}{   6}
\z{1073}{ 626}{   1}\x{1073}{ 625}{   6}
\z{1079}{ 625}{   2}\x{1079}{ 623}{   6}
\z{1085}{ 623}{   1}\x{1085}{ 622}{   6}
\z{1091}{ 622}{   1}\x{1091}{ 621}{   6}
\z{1097}{ 621}{   1}\x{1097}{ 620}{   6}
\z{1103}{ 620}{   1}\x{1103}{ 619}{   6}
\z{1109}{ 619}{   1}\x{1109}{ 618}{   6}
\z{1115}{ 618}{   2}\x{1115}{ 616}{   6}
\z{1121}{ 616}{   1}\x{1121}{ 615}{   6}
\z{1127}{ 615}{   1}\x{1127}{ 614}{   6}
\z{1133}{ 614}{   1}\x{1133}{ 613}{   6}
\z{1139}{ 613}{   2}\x{1139}{ 611}{   6}
\z{1145}{ 611}{   1}\x{1145}{ 610}{   6}
\z{1151}{ 610}{   1}\x{1151}{ 609}{   6}
\z{1157}{ 609}{   1}\x{1157}{ 608}{   6}
\z{1163}{ 608}{   2}\x{1163}{ 606}{   6}
\z{1169}{ 606}{   1}\x{1169}{ 605}{   6}
\z{1175}{ 605}{   1}\x{1175}{ 604}{   6}
\z{1181}{ 604}{   1}\x{1181}{ 603}{   6}
\z{1187}{ 603}{   2}\x{1187}{ 601}{   6}
\z{1193}{ 601}{   1}\x{1193}{ 600}{   6}
\end{picture}} 
\end{picture} 
\caption{ \baselineskip=7mm
Direct CP asymmetry for $\bar B\rightarrow \pi^+\pi^-$,
${\cal A}^{dir}_{CP}(\pi\pi)_d$,
for $R_b=0.385$ as
calculated from the amplitude in Eq.(5) of Ref.~\cite{pipi},
which is derived from the analoga of the diagrams in Figs.~1 - 3.}
\end{figure}
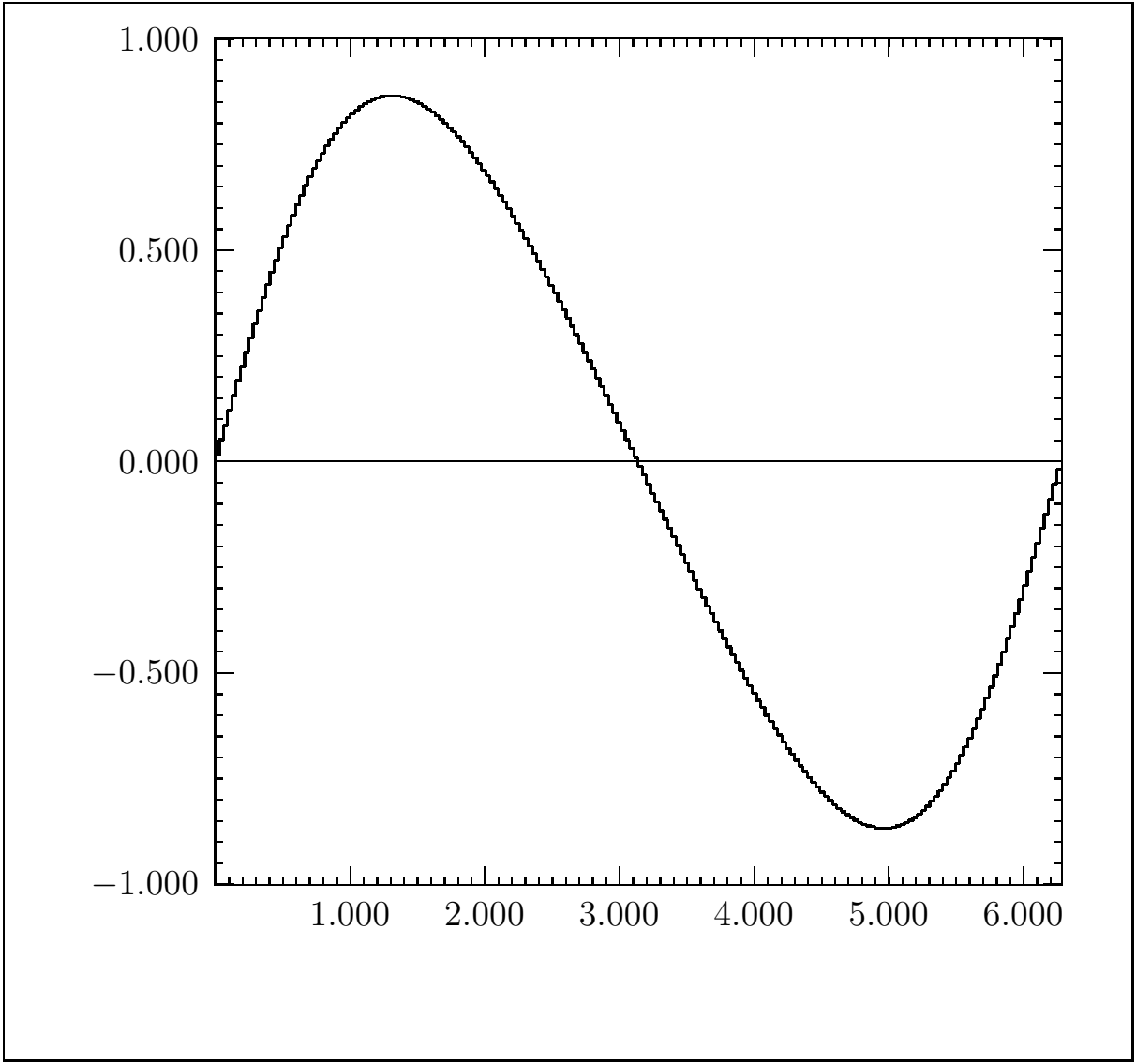
\begin{figure} 
\begin{center}
  Direct CP Asymmetry ${\cal A}^{dir}_{CP}(\rho K_S)_s$                        \end{center}
\setlength{\unitlength}{0.1mm}
\begin{picture}(1600,1500)
\put(0,0){\framebox(1600,1500){ }}
\put(300,250){\begin{picture}( 1200,1200)
\put(0,0){\framebox( 1200,1200){ }}
\multiput(  190.99,0)(  190.99,0){   6}{\line(0,1){25}}
\multiput(     .00,0)(   19.10,0){  63}{\line(0,1){10}}
\multiput(  190.99,1200)(  190.99,0){   6}{\line(0,-1){25}}
\multiput(     .00,1200)(   19.10,0){  63}{\line(0,-1){10}}
\put( 190,-25){\makebox(0,0)[t]{\large $    1.000 $}}
\put( 381,-25){\makebox(0,0)[t]{\large $    2.000 $}}
\put( 572,-25){\makebox(0,0)[t]{\large $    3.000 $}}
\put( 763,-25){\makebox(0,0)[t]{\large $    4.000 $}}
\put( 954,-25){\makebox(0,0)[t]{\large $    5.000 $}}
\put(1145,-25){\makebox(0,0)[t]{\large $    6.000 $}}
\multiput(0,     .00)(0,  300.00){   5}{\line(1,0){25}}
\multiput(0,     .00)(0,   30.00){  41}{\line(1,0){10}}
\multiput(1200,     .00)(0,  300.00){   5}{\line(-1,0){25}}
\multiput(1200,     .00)(0,   30.00){  41}{\line(-1,0){10}}
\put(0,  600.00){\line(1,0){1200}}
\put(-25,   0){\makebox(0,0)[r]{\large $  -1.000 $}}
\put(-25, 300){\makebox(0,0)[r]{\large $   -0.500 $}}
\put(-25, 600){\makebox(0,0)[r]{\large $     0.000 $}}
\put(-25, 900){\makebox(0,0)[r]{\large $    0.500 $}}
\put(-25,1200){\makebox(0,0)[r]{\large $   1.000 $}}
\end{picture}}
\put(300,250){\begin{picture}( 1200,1200)
\thicklines 
\newcommand{\x}[3]{\put(#1,#2){\line(1,0){#3}}}
\newcommand{\y}[3]{\put(#1,#2){\line(0,1){#3}}}
\newcommand{\z}[3]{\put(#1,#2){\line(0,-1){#3}}}
\newcommand{\e}[3]{\put(#1,#2){\line(0,1){#3}}}
\y{   0}{   0}{ 610}\x{   0}{ 610}{   5}
\y{   5}{ 610}{  21}\x{   5}{ 631}{   6}
\y{  11}{ 631}{  21}\x{  11}{ 652}{   6}
\y{  17}{ 652}{  21}\x{  17}{ 673}{   6}
\y{  23}{ 673}{  21}\x{  23}{ 694}{   6}
\y{  29}{ 694}{  21}\x{  29}{ 715}{   6}
\y{  35}{ 715}{  20}\x{  35}{ 735}{   6}
\y{  41}{ 735}{  20}\x{  41}{ 755}{   6}
\y{  47}{ 755}{  20}\x{  47}{ 775}{   6}
\y{  53}{ 775}{  20}\x{  53}{ 795}{   6}
\y{  59}{ 795}{  19}\x{  59}{ 814}{   6}
\y{  65}{ 814}{  19}\x{  65}{ 833}{   6}
\y{  71}{ 833}{  18}\x{  71}{ 851}{   6}
\y{  77}{ 851}{  18}\x{  77}{ 869}{   6}
\y{  83}{ 869}{  17}\x{  83}{ 886}{   6}
\y{  89}{ 886}{  17}\x{  89}{ 903}{   6}
\y{  95}{ 903}{  16}\x{  95}{ 919}{   6}
\y{ 101}{ 919}{  16}\x{ 101}{ 935}{   6}
\y{ 107}{ 935}{  15}\x{ 107}{ 950}{   6}
\y{ 113}{ 950}{  14}\x{ 113}{ 964}{   6}
\y{ 119}{ 964}{  14}\x{ 119}{ 978}{   6}
\y{ 125}{ 978}{  14}\x{ 125}{ 992}{   6}
\y{ 131}{ 992}{  12}\x{ 131}{1004}{   6}
\y{ 137}{1004}{  12}\x{ 137}{1016}{   6}
\y{ 143}{1016}{  11}\x{ 143}{1027}{   6}
\y{ 149}{1027}{  11}\x{ 149}{1038}{   6}
\y{ 155}{1038}{  10}\x{ 155}{1048}{   6}
\y{ 161}{1048}{   9}\x{ 161}{1057}{   6}
\y{ 167}{1057}{   9}\x{ 167}{1066}{   6}
\y{ 173}{1066}{   8}\x{ 173}{1074}{   6}
\y{ 179}{1074}{   7}\x{ 179}{1081}{   6}
\y{ 185}{1081}{   7}\x{ 185}{1088}{   6}
\y{ 191}{1088}{   6}\x{ 191}{1094}{   6}
\y{ 197}{1094}{   5}\x{ 197}{1099}{   6}
\y{ 203}{1099}{   5}\x{ 203}{1104}{   6}
\y{ 209}{1104}{   4}\x{ 209}{1108}{   6}
\y{ 215}{1108}{   3}\x{ 215}{1111}{   6}
\y{ 221}{1111}{   3}\x{ 221}{1114}{   6}
\y{ 227}{1114}{   2}\x{ 227}{1116}{   6}
\y{ 233}{1116}{   2}\x{ 233}{1118}{   6}
\y{ 239}{1118}{   1}\x{ 239}{1119}{   6}
\y{ 245}{1119}{   0}\x{ 245}{1119}{   6}
\y{ 251}{1119}{   0}\x{ 251}{1119}{   6}
\y{ 257}{1119}{   0}\x{ 257}{1119}{   6}
\z{ 263}{1119}{   2}\x{ 263}{1117}{   6}
\z{ 269}{1117}{   1}\x{ 269}{1116}{   6}
\z{ 275}{1116}{   2}\x{ 275}{1114}{   6}
\z{ 281}{1114}{   3}\x{ 281}{1111}{   6}
\z{ 287}{1111}{   3}\x{ 287}{1108}{   6}
\z{ 293}{1108}{   4}\x{ 293}{1104}{   6}
\z{ 299}{1104}{   4}\x{ 299}{1100}{   6}
\z{ 305}{1100}{   4}\x{ 305}{1096}{   6}
\z{ 311}{1096}{   5}\x{ 311}{1091}{   6}
\z{ 317}{1091}{   5}\x{ 317}{1086}{   6}
\z{ 323}{1086}{   6}\x{ 323}{1080}{   6}
\z{ 329}{1080}{   6}\x{ 329}{1074}{   6}
\z{ 335}{1074}{   6}\x{ 335}{1068}{   6}
\z{ 341}{1068}{   7}\x{ 341}{1061}{   6}
\z{ 347}{1061}{   7}\x{ 347}{1054}{   6}
\z{ 353}{1054}{   7}\x{ 353}{1047}{   6}
\z{ 359}{1047}{   8}\x{ 359}{1039}{   6}
\z{ 365}{1039}{   8}\x{ 365}{1031}{   6}
\z{ 371}{1031}{   8}\x{ 371}{1023}{   6}
\z{ 377}{1023}{   9}\x{ 377}{1014}{   6}
\z{ 383}{1014}{   8}\x{ 383}{1006}{   6}
\z{ 389}{1006}{   9}\x{ 389}{ 997}{   6}
\z{ 395}{ 997}{  10}\x{ 395}{ 987}{   6}
\z{ 401}{ 987}{   9}\x{ 401}{ 978}{   6}
\z{ 407}{ 978}{  10}\x{ 407}{ 968}{   6}
\z{ 413}{ 968}{   9}\x{ 413}{ 959}{   6}
\z{ 419}{ 959}{  11}\x{ 419}{ 948}{   6}
\z{ 425}{ 948}{  10}\x{ 425}{ 938}{   6}
\z{ 431}{ 938}{  10}\x{ 431}{ 928}{   6}
\z{ 437}{ 928}{  11}\x{ 437}{ 917}{   6}
\z{ 443}{ 917}{  11}\x{ 443}{ 906}{   6}
\z{ 449}{ 906}{  11}\x{ 449}{ 895}{   6}
\z{ 455}{ 895}{  11}\x{ 455}{ 884}{   6}
\z{ 461}{ 884}{  11}\x{ 461}{ 873}{   6}
\z{ 467}{ 873}{  11}\x{ 467}{ 862}{   6}
\z{ 473}{ 862}{  12}\x{ 473}{ 850}{   6}
\z{ 479}{ 850}{  11}\x{ 479}{ 839}{   6}
\z{ 485}{ 839}{  12}\x{ 485}{ 827}{   6}
\z{ 491}{ 827}{  12}\x{ 491}{ 815}{   6}
\z{ 497}{ 815}{  11}\x{ 497}{ 804}{   6}
\z{ 503}{ 804}{  12}\x{ 503}{ 792}{   6}
\z{ 509}{ 792}{  12}\x{ 509}{ 780}{   6}
\z{ 515}{ 780}{  12}\x{ 515}{ 768}{   6}
\z{ 521}{ 768}{  13}\x{ 521}{ 755}{   6}
\z{ 527}{ 755}{  12}\x{ 527}{ 743}{   6}
\z{ 533}{ 743}{  12}\x{ 533}{ 731}{   6}
\z{ 539}{ 731}{  13}\x{ 539}{ 718}{   6}
\z{ 545}{ 718}{  12}\x{ 545}{ 706}{   6}
\z{ 551}{ 706}{  12}\x{ 551}{ 694}{   6}
\z{ 557}{ 694}{  13}\x{ 557}{ 681}{   6}
\z{ 563}{ 681}{  12}\x{ 563}{ 669}{   6}
\z{ 569}{ 669}{  13}\x{ 569}{ 656}{   6}
\z{ 575}{ 656}{  12}\x{ 575}{ 644}{   6}
\z{ 581}{ 644}{  13}\x{ 581}{ 631}{   6}
\z{ 587}{ 631}{  13}\x{ 587}{ 618}{   6}
\z{ 593}{ 618}{  12}\x{ 593}{ 606}{   6}
\z{ 599}{ 606}{  13}\x{ 599}{ 593}{   6}
\z{ 605}{ 593}{  12}\x{ 605}{ 581}{   6}
\z{ 611}{ 581}{  13}\x{ 611}{ 568}{   6}
\z{ 617}{ 568}{  13}\x{ 617}{ 555}{   6}
\z{ 623}{ 555}{  12}\x{ 623}{ 543}{   6}
\z{ 629}{ 543}{  13}\x{ 629}{ 530}{   6}
\z{ 635}{ 530}{  12}\x{ 635}{ 518}{   6}
\z{ 641}{ 518}{  13}\x{ 641}{ 505}{   6}
\z{ 647}{ 505}{  12}\x{ 647}{ 493}{   6}
\z{ 653}{ 493}{  12}\x{ 653}{ 481}{   6}
\z{ 659}{ 481}{  13}\x{ 659}{ 468}{   6}
\z{ 665}{ 468}{  12}\x{ 665}{ 456}{   6}
\z{ 671}{ 456}{  12}\x{ 671}{ 444}{   6}
\z{ 677}{ 444}{  13}\x{ 677}{ 431}{   6}
\z{ 683}{ 431}{  12}\x{ 683}{ 419}{   6}
\z{ 689}{ 419}{  12}\x{ 689}{ 407}{   6}
\z{ 695}{ 407}{  12}\x{ 695}{ 395}{   6}
\z{ 701}{ 395}{  11}\x{ 701}{ 384}{   6}
\z{ 707}{ 384}{  12}\x{ 707}{ 372}{   6}
\z{ 713}{ 372}{  12}\x{ 713}{ 360}{   6}
\z{ 719}{ 360}{  11}\x{ 719}{ 349}{   6}
\z{ 725}{ 349}{  12}\x{ 725}{ 337}{   6}
\z{ 731}{ 337}{  11}\x{ 731}{ 326}{   6}
\z{ 737}{ 326}{  11}\x{ 737}{ 315}{   6}
\z{ 743}{ 315}{  11}\x{ 743}{ 304}{   6}
\z{ 749}{ 304}{  11}\x{ 749}{ 293}{   6}
\z{ 755}{ 293}{  11}\x{ 755}{ 282}{   6}
\z{ 761}{ 282}{  11}\x{ 761}{ 271}{   6}
\z{ 767}{ 271}{  10}\x{ 767}{ 261}{   6}
\z{ 773}{ 261}{  10}\x{ 773}{ 251}{   6}
\z{ 779}{ 251}{  11}\x{ 779}{ 240}{   6}
\z{ 785}{ 240}{   9}\x{ 785}{ 231}{   6}
\z{ 791}{ 231}{  10}\x{ 791}{ 221}{   6}
\z{ 797}{ 221}{   9}\x{ 797}{ 212}{   6}
\z{ 803}{ 212}{  10}\x{ 803}{ 202}{   6}
\z{ 809}{ 202}{   9}\x{ 809}{ 193}{   6}
\z{ 815}{ 193}{   8}\x{ 815}{ 185}{   6}
\z{ 821}{ 185}{   9}\x{ 821}{ 176}{   6}
\z{ 827}{ 176}{   8}\x{ 827}{ 168}{   6}
\z{ 833}{ 168}{   8}\x{ 833}{ 160}{   6}
\z{ 839}{ 160}{   8}\x{ 839}{ 152}{   6}
\z{ 845}{ 152}{   7}\x{ 845}{ 145}{   6}
\z{ 851}{ 145}{   7}\x{ 851}{ 138}{   6}
\z{ 857}{ 138}{   7}\x{ 857}{ 131}{   6}
\z{ 863}{ 131}{   6}\x{ 863}{ 125}{   6}
\z{ 869}{ 125}{   6}\x{ 869}{ 119}{   6}
\z{ 875}{ 119}{   6}\x{ 875}{ 113}{   6}
\z{ 881}{ 113}{   5}\x{ 881}{ 108}{   6}
\z{ 887}{ 108}{   5}\x{ 887}{ 103}{   6}
\z{ 893}{ 103}{   4}\x{ 893}{  99}{   6}
\z{ 899}{  99}{   4}\x{ 899}{  95}{   6}
\z{ 905}{  95}{   4}\x{ 905}{  91}{   6}
\z{ 911}{  91}{   3}\x{ 911}{  88}{   6}
\z{ 917}{  88}{   3}\x{ 917}{  85}{   6}
\z{ 923}{  85}{   2}\x{ 923}{  83}{   6}
\z{ 929}{  83}{   1}\x{ 929}{  82}{   6}
\z{ 935}{  82}{   2}\x{ 935}{  80}{   6}
\y{ 941}{  80}{   0}\x{ 941}{  80}{   6}
\y{ 947}{  80}{   0}\x{ 947}{  80}{   6}
\y{ 953}{  80}{   0}\x{ 953}{  80}{   6}
\y{ 959}{  80}{   1}\x{ 959}{  81}{   6}
\y{ 965}{  81}{   2}\x{ 965}{  83}{   6}
\y{ 971}{  83}{   2}\x{ 971}{  85}{   6}
\y{ 977}{  85}{   3}\x{ 977}{  88}{   6}
\y{ 983}{  88}{   3}\x{ 983}{  91}{   6}
\y{ 989}{  91}{   4}\x{ 989}{  95}{   6}
\y{ 995}{  95}{   5}\x{ 995}{ 100}{   6}
\y{1001}{ 100}{   5}\x{1001}{ 105}{   6}
\y{1007}{ 105}{   6}\x{1007}{ 111}{   6}
\y{1013}{ 111}{   7}\x{1013}{ 118}{   6}
\y{1019}{ 118}{   7}\x{1019}{ 125}{   6}
\y{1025}{ 125}{   8}\x{1025}{ 133}{   6}
\y{1031}{ 133}{   9}\x{1031}{ 142}{   6}
\y{1037}{ 142}{   9}\x{1037}{ 151}{   6}
\y{1043}{ 151}{  10}\x{1043}{ 161}{   6}
\y{1049}{ 161}{  11}\x{1049}{ 172}{   6}
\y{1055}{ 172}{  11}\x{1055}{ 183}{   6}
\y{1061}{ 183}{  12}\x{1061}{ 195}{   6}
\y{1067}{ 195}{  12}\x{1067}{ 207}{   6}
\y{1073}{ 207}{  14}\x{1073}{ 221}{   6}
\y{1079}{ 221}{  14}\x{1079}{ 235}{   6}
\y{1085}{ 235}{  14}\x{1085}{ 249}{   6}
\y{1091}{ 249}{  15}\x{1091}{ 264}{   6}
\y{1097}{ 264}{  16}\x{1097}{ 280}{   6}
\y{1103}{ 280}{  16}\x{1103}{ 296}{   6}
\y{1109}{ 296}{  17}\x{1109}{ 313}{   6}
\y{1115}{ 313}{  17}\x{1115}{ 330}{   6}
\y{1121}{ 330}{  18}\x{1121}{ 348}{   6}
\y{1127}{ 348}{  18}\x{1127}{ 366}{   6}
\y{1133}{ 366}{  19}\x{1133}{ 385}{   6}
\y{1139}{ 385}{  19}\x{1139}{ 404}{   6}
\y{1145}{ 404}{  20}\x{1145}{ 424}{   6}
\y{1151}{ 424}{  20}\x{1151}{ 444}{   6}
\y{1157}{ 444}{  20}\x{1157}{ 464}{   6}
\y{1163}{ 464}{  20}\x{1163}{ 484}{   6}
\y{1169}{ 484}{  21}\x{1169}{ 505}{   6}
\y{1175}{ 505}{  21}\x{1175}{ 526}{   6}
\y{1181}{ 526}{  21}\x{1181}{ 547}{   6}
\y{1187}{ 547}{  21}\x{1187}{ 568}{   6}
\y{1193}{ 568}{  21}\x{1193}{ 589}{   6}
\end{picture}} 
\end{picture} 
\caption{ \baselineskip=7mm
Direct CP asymmetry for $\bar B_s\rightarrow \rho K_s$,
${\cal A}^{dir}_{CP}(\rho K_S)_s$, 
for $R_b=0.385$ as
calculated from the diagrams in Figs.~1 - 3.}
\end{figure}
This should be compared to the
result of Beneke {et al.}~\cite{beneke}, $-0.04\times \sin\gamma$.
Evidently, experiment will soon distinguish these two results.
For reference, we also record the direct CP violating asymmetry
for the $\bar B_s\rightarrow \rho K_S$ case, ${\cal A}^{dir}_{CP}(\rho K_S)_s$,
as a function of $\gamma$ in Fig.~6.
We see that it is substantial in a large part of the
preferred regime $45^o\le \gamma \le 135^o$, just as
it is a large part of its nonzero value in this region in
the case of ${\cal A}^{dir}_{CP}(\pi^+\pi^-)_d$.
The recoil phase effect is an essential part of the results
in Figs. 5 and 6.
For proving CP violation in the B system, these modes
suggest that a measurement of ${\cal A}^{dir}_{CP}$ may be a reasonable
way to proceed.\par
Next, we turn to the case of the modes $D^*\pi$, where we
follow the notation of Ref.~\cite{th-2000-101} and refer
to $f=D^{*+}\pi^-,~\bar f=D^{*-}\pi^+$. A strategy advocated in
Ref.~\cite{th-2000-101} is to measure the combination
$2\beta+\gamma$ in the time-dependent asymmetries
for $\bar B\rightarrow f$ and $\bar B\rightarrow \bar f$
using the fact that the product $\xi_f^{(d)}\times \xi_{\bar f}^{(d)}$
yields $e^{-2i(2\beta+\gamma)}$ if we define (here, $\phi_d$ is the $B_d$
mixing phase $2\beta$, $\lambda \equiv|V_{us}|$ )
\begin{equation}
\xi^{(d)}_f =-e^{-i\phi_d}\frac{A(\bar{B}^0_d\rightarrow f)}{A(B^0_d\rightarrow f)}=-e^{-i(\phi_d+\gamma)}\left(\frac{1-\lambda^2}{\lambda^2R_b}\right)
\frac{\bar M_f}{M_{\bar f}}\cr
\xi^{(d)}_{\bar f} =-e^{-i\phi_d}\frac{A(\bar{B}^0_d\rightarrow\bar f)}{A(B^0_d\rightarrow \bar f)}=-e^{-i(\phi_d+\gamma)}
\left(\frac{\lambda^2R_b}{1-\lambda^2}\right)\frac{ M_{\bar f}}{\bar M_{f}},
\label{eq:cptime}
\end{equation}
for the amplitudes $A(\bar{B}^0_d\rightarrow f,\bar f)$ and their CP conjugates
respectively. Thus, $\bar M_f,~ M_{\bar f}$ are the respective strong
interaction matrix elements defined in Eq.(3.26) of Ref.~\cite{th-2000-101}.
The point is that, in
the actual extraction of the time dependent
asymmetry, the strong recoil phase effect gives a non-trivial
value to the strong phase $\Delta_S$, as 
defined in Ref.~\cite{th-2000-101}, 
in the ratio $\bar M_f/M_{\bar f}$. In Ref.~\cite{th-2000-101},
this phase has been set to 0 to estimate how accurately
the weak phase could be measured in the LHCB environment.
Upon calculating the analogue of Fig. 1 for these processes,
we find that the value of $\Delta_S$ is $-253.6^o$. Thus,
the analysis in Ref.~\cite{th-2000-101} should address
non-trivial values of $\Delta_S$ also.
\par
The analysis in Ref.~\cite{th-2000-101} also attempts
to use u-spin and SU(3) symmetry to isolate $\gamma$
in several modes, $\bar B\rightarrow \pi K$, $\bar B_{s,d}\rightarrow \Psi/J K_S$,and  $\bar B\rightarrow \pi \pi,~K K$ modes. 
Here, we discuss the perturbative QCD
expectations for these assumptions. Since the tree and penguin contributions
enter with different CKM coefficients, $V_{\cal UD}^*V_{{\cal U}b}$,
to show the inadequacy of u-spin symmetry, it is enough to
focus on the analogue of Fig.~1 for these decays. The complete predictions
from the analogue of all the graphs in Figs.~1-3 will appear 
elsewhere~\cite{elsewhere}. For the processes $\bar B_{s,d} ->\Psi/J K_S$
we find for the analoga of Fig. 1 the recoil phases
\begin{equation}
\delta_T(B_s)=0.982,~ \delta_T(B_d)=2.24, 
\label{recph1}
\end{equation}
and the ratio of strong transition amplitude moduli squared
\begin{equation}
|{\cal A'}|^2/|{\cal A}|^2= 1.81,
\label{recph2}
\end{equation}
where $\delta_T(B_s),~ \delta_T{B_d}$ are the respective
strong recoil phases for the graphs in Fig.~1 for the $\bar B_d$
and $\bar B_s$ cases respectively and ${\cal A}',~{\cal A}$ are
the respective strong transition amplitudes. Evidently,
the assumption of SU(3) and u-spin symmetry in exclusive
B decays to light mesons is completely unfounded
and the recoil phase effect makes the  situation even more acute;
for, if the recoil phase is ignored, the 1.81 in (\ref{recph2})
becomes 2.24.
\par
In summary, we have shown that the physical phenomenon
of the recoil phase effect is important for CP violation
studies in $B$ decays to two light mesons. We have shown how
to take it into account in Refs.~\cite{semdcy,dppkk,pipi,rks1}.
We look forward to its further application to the exciting 
field of CP violation studies in exclusive $B$ decays.\par
{\large\bf  Acknowledgements}
 
The author acknowledges the
kind hospitality of Prof. C. Prescott and SLAC Group A
and helpful discussions with Drs. P. Dauncey and Robert Fleischer
and Prof. L. Lanceri at various stages of this work.
The author thanks Prof. M. Jezabek and the Organizing Committee
for inviting him to lecture in the 2001 Cracow Epiphany Conference.\\
Notes Added:\\
1. The imaginary parts which we find in the recoil exchanges
in Figs. 1-3 are all leading twist effects. They arise from
the (anomalous) solutions of the respective Cutkowsky-Landau-Bjorken
equations associated with these graphs, as described in the book by
J. D. Bjorken and S. D. Drell, \underline{ Relativistic Quantum Fields},
(McGraw-Hill, Menlo Park, 1965). Any consistent dispersive treatment
of these graphs has to take all of these solutions into account,
both anomalous and non-anomalous solutions.\\
2. As the semi-leptonic decay distribution has the form
$d\Gamma(B\rightarrow X_{\cal U}+\ell+\nu_\ell)=|V_{{\cal U}b}|^2
|F^{\cal U}_{QCD}|^2 dLIPS$,~${\cal U}=u,c$, where $dLIPS$
is the respective Lorentz invariant phase space factor and
both the moduli $|V_{{\cal U}b}|$ and the strong interaction transition
amplitude factor $F^{\cal U}_{QCD}$ are CP invariant, it follows that
the analogue of the recoil phase in Fig. 1 for the semi-leptonic
decays does not generate CP violation in these decays.\\
3. We finally stress that the Lepage-Brodsky expansion in Ref.~\cite{L-B}
is an exact re-arrangement of the exact Bethe-Salpeter bound state
transition amplitude. Only when authors make arbitrary truncations
of the expansion, for example, treating the endpoint contributions
at higher twist without including the respective Sudakov resummation
that makes them finite, do unknown parameters appear in the
application of the expansion to hard interaction processes such as
exclusive B decays to two light mesons.

\end{document}